\documentclass{acm_proc_article-sp}

\begin{document}

\title{Internet topology at the router and  autonomous system level}

\numberofauthors{3}
\author{
\alignauthor {Alexei V{\'a}zquez}\\
       \affaddr{\small{INFM and International School for Advanced Studies, 
via Beirut 4, 34014 Trieste, Italy}}
%       \affaddr{via Beirut 4}\\
%       \affaddr{34014 Trieste, Italy}\\
       \email{\small{vazquez@sissa.it}}
\alignauthor \mbox{Romualdo Pastor-Satorras}\\
        \affaddr{\small{Dept. de F{\'\i}sica i Enginyeria Nuclear, 
                   Universitat Polit{\`e}cnica de Catalunya 
Campus Nord, M\`{o}dul B4, 08034 Barcelona, Spain}}
        \email{\small{romu@sinera.upc.es}}
\alignauthor {Alessandro Vespignani}\\
        \affaddr{\small{The Abdus Salam International 
                 Centre for Theoretical Physics, P.O. Box 586, 
          34100 Trieste, Italy}}
        \email{\small{alexv@ictp.trieste.it}}
}
\maketitle
\date{\today}

\begin{abstract}

  We present a statistical analysis of different metrics
  characterizing the topological properties of Internet maps,
  collected at two different resolution scales: the router  and
  the autonomous system level. The metrics we consider allow us to
  confirm the presence of scale-free signatures in several statistical
  distributions, as well as to show in a quantitative way the
  hierarchical nature of the Internet. Our findings are relevant for
  the development of more accurate Internet topology generators, which
  should include, along with the scale-free properties
  of the connectivity distribution, the hierarchical signatures unveiled
  in the present work.

\end{abstract}

\section{Introduction}

The relentless growth of the Internet goes along with a wide range of
internetworking problems related to routing protocols, resource
allowances, and physical connectivity plans. The study and
optimization of algorithms and policies related to such problems rely
heavily on theoretical analysis and simulations that use model
abstractions of the actual Internet.  On the other hand, in order to
extract the maximum benefit from these studies, it is necessary to
work with reliable Internet topology generators. 
The basic priority at this respect is to best define the topology 
to use for the network being simulated.
This implies the characterization of how routers, hosts, and physical
links interconnect with each other in shaping the actual Internet.

In the last years, research groups started to deploy technologies and
infrastructures in order to obtain a more detailed picture of the
Internet. Several studies, aimed at tracking and visualizing the
Internet large scale topology and/or performance, are leading to
Internet mapping projects at different resolution scales. These
projects typically collect data on Internet elements (routers,
domains) and the connections among them (physical links, peer
connections), in order to create a graph-like representation of large
parts of the Internet in which the nodes represent those elements and
the links represent the respective connections.  Mapping projects
focus essentially on two levels of topological description.  First, by
inferring router adjacencies it has been possible to measure the
Internet router (IR) level topology.  The second measured topology
works at the autonomous system (AS) level and the connectivity
obtained from AS routing path information.  Although these two
representations are related, it is clear that they describe the
Internet at rather different scales. In fact, each AS groups a
generally large number of routers, and therefore the AS maps are in
some sense a coarse-grained view of the IR maps.

Internet maps  exhibit an extremely large degree of heterogeneity and
the use of statistical tools becomes mandatory to provide a proper
mathematical characterization of this system.  Statistical analysis of
the Internet maps fabric have pointed out, to the surprise of many
researchers, a very complex connectivity pattern with fluctuations
extending over several orders of magnitude \cite{faloutsos}.  In
particular, it has been observed a power-law behavior in
metrics and statistical distributions of Internet maps at different
levels
\cite{faloutsos,mercator,calda,us02,willinger02,tangmu02,chen02,broido,erl}.
This evidence makes the Internet an example of the so-called
\textit{scale-free} networks \cite{albert02} and uncover a peculiar
structure that cannot be satisfactorily modeled with traditional
topology generators.  Previous Internet topology generators, based in
the classical Erd{\"o}s and R{\'e}nyi random graph model \cite{er,bollobas} or
in hierarchical models, yielded an exponentially bounded connectivity
pattern, with very small fluctuations and in clear disagreement with
the recent empirical findings.  A theoretical framework for the origin
of scale-free graphs has been put forward by Barab\'{a}si and Albert
\cite{albert02} by devising a novel class of dynamical growing
networks.  Following these ideas, several Internet topology generators
yielding power-law distributions have been subsequently proposed
\cite{medina,brite,inet}.

Data gathering projects \cite{nlanr,caida,as_data,scan,lucent} are
progressively making available larger AS and IR level maps which are
susceptible of more accurate statistical analysis and raise new and
challenging questions about the Internet topology.  For instance,
statistical distributions show deviations from the pure power-law
behavior and it is important to understand to which extent the
Internet can be considered a scale-free graph. The way these scaling
anomalies---usually signaled by the presence of cut-offs in the
corresponding statistical distributions---are related to the Internet
finite size and physical constraints is a capital issue in the
characterization of the Internet and in the understanding of the
dynamics underlying its growth.  A further important issue
concerns the fact that the Internet is  organized on different
hierarchical levels, with a set of backbone links carrying the traffic
between local area providers.  This structure is reflected in a
hierarchical arrangement of administrative domains and in a different
usage of links and connectivity of nodes. The interplay between the
scale-free nature and the hierarchical properties of the Internet is
still unclear, and it is an important task to find metrics that can
exploit and characterize hierarchical features on the AS and IR level.
Finally, although one would expect Internet AS and IR level maps to
exhibit similar scale-free properties, the different resolution in
both kinds of maps might lead to a diversity of metrics properties.

In this paper we present a detailed statistical analysis of large AS
and IR level maps \cite{nlanr,as_data,scan}.  We study the scale-free
properties of these maps, focusing on the degree and betweenness
distributions. While scale-free properties are confirmed for maps at
both levels, IR level maps show also the presence of an exponential
cut-off, that can be related to constraints acting on the physical
connectivity and load of routers. Power-law distributions with a
cut-off are a general feature of scale-free phenomena in real finite
systems and we discuss their origin in the framework of growing
networks.  At the AS level we confirm the presence of a strong
scale-free character for the large-scale degree and betweenness
distributions.  We also discuss that deviations from the pure
power-law behavior found in recent maps \cite{as_data} at intermediate
connectivities has a marginal impact on the resilience and information
spreading properties of the Internet \cite{resilience,spreading}.

Furthermore, we propose two metrics based on the connectivity and the
clustering correlation functions, that appear to sharply characterize
the hierarchical properties of Internet maps. In particular, these
metrics clearly distinguish between the AS and IR levels, which show a
very different behavior at this respect. While IR level maps appear to
possess almost no hierarchical structure, AS maps fully exploit the
hierarchy of domains around which the Internet revolves.  The
differences highlighted between the two levels might be very important
in the developing of faithful Internet topology generators.  The
testing of Internet protocols working at different levels might need
of topology generators accounting for the different properties
observed.  Hierarchical features are also important to scrutinize
theoretical models proposing new dynamical growth mechanisms for the
Internet as a whole.

\section{Internet maps}

Nowadays the Internet can be partitioned in
autonomously administered domains which vary in size, geographical
extent, and function. Each domain may exercise traffic restrictions or
preferences, and handle internal traffic according to particular
autonomous policies.  This fact has stimulated the separation of the
inter-domain routing from the intra-domain routing, and the
introduction of the Autonomous Systems Number (ASN). Each AS refers to
one single administrative domain of the Internet. Within each AS, an
Interior Gateway Protocol is used for routing purposes. Between ASs,
an Exterior Gateway Protocol provides the inter-domain routing system.
The Border Gateway Protocol (BGP) is the most widely used inter-domain
protocol.  In particular, it assigns a 16-bit ASN to identify, and
refer to, each AS.

The Internet is usually portrayed as an undirected graph. Depending on
the meaning assigned to the nodes and links of the associated graph,
we can obtain different levels of representation, each one
corresponding to a different degree of coarse-graining respect to the
physical Internet.

{\em Internet Router level}: In the IR level maps, nodes represents
the routers, while links represent the physical connections among them.
In general, all mapping efforts at the IR level are based on computing
router adjacencies from {\em traceroute} sequences sent to a list of
networks in the Internet. The traceroute command performed from a
single source provides a spanning tree from that source to every other
(reachable) node in the network. By merging the information obtained
from different sources it is possible to construct IR level maps of
different portions of the Internet. In order to catch all the various
cross-links, however, a large number of source probes is needed.  In
addition, the instability of paths  between routers and other
technical problems---such as multiple alias interfaces---make the
mapping a very difficult task \cite{burch99}.  These difficulties have
been diversely tackled by the different Internet mapping projects: the
Lucent project at Bell Labs \cite{lucent}, the Cooperative Association
for Internet Data Analysis \cite{caida}, and the SCAN project at the
Information Sciences Institute \cite{scan}, that develop methods to
obtain partial maps from a single source.

{\em Autonomous System level}: In the AS level graphs each node
represents an AS, while each link between two nodes represents the
existence of a BGP peer connection among the corresponding ASs.  It is
important to stress that each AS groups many routers together and the
traffic carried by a link is the aggregation of all the individual
end-host flows between the corresponding ASs.  The AS map can be
constructed by looking at the BGP routing tables. In fact, the BGP
routing tables of each AS contains a spanning tree from that node to
every other (reachable) AS. We can then try to reconstruct the
complete AS map by merging the connectivity information coming from a
certain fraction of these spanning trees. This method has been
actually used by the National Laboratory for Applied Network Research
(NLANR) \cite{nlanr}, using the BGP routing tables collected at the
Oregon route server, that gathers BGP-related information since 1997.
Enriched maps can be obtained from some other public sources, such as
Looking Glass sites and the Reseaux IP Europeens (RIPE) \cite{chen02},
getting about 40\% of new AS-AS connections.

These graph representations do not model individual hosts, too
numerous, and neglect link properties such as bandwidth, actual data
load, or geographical distance. For these reasons, the graph-like
representation must be considered as an overlay of the basic
topological structure: the skeleton of the Internet.  Moreover, the
data collected for the two levels are different, and both
representations may be incomplete or partial to different degrees.  In
particular, measurements may not capture all the nodes present in the
actual network and, more often, they do not include all the links
among nodes. It is not our purpose here to argue about the reliability
of the different maps.  However, the conclusions we shall present in
this paper seem rather stable in time for the different maps.
Hopefully, this fact means that, despite the different degrees of
completeness, the present maps represent a fairly good statistical
sampling of the Internet as a whole.  In particular, we shall use the
map collected during October/November 1999 by the SCAN project with
the Mercator software as representative of the Internet router level.
At the autonomous system level we consider the (AS) map collected at
Oregon route server and the enriched (AS+) map (available at
\cite{as_data}), both dated May 25, 2001.

\section{Average properties}
\label{sec:ave}

We start our study by analysing some standard
metrics: the total number of nodes $N$ and edges $E$, the node
connectivity $k_i$, the minimum path distance between pairs of nodes
$d_{ij}$, the clustering coefficient $c_i$, and the betweenness $b_i$.
The connectivity $k_i$ of a node is defined as the number of edges
incident to that node, {\em i.e.}  the number of connections of that
node with other nodes in the network.  If nodes $i$ and $j$ are
connected we will say that they are nearest neighbors. The minimum
path distance $d_{ij}$ between a pair of nodes $i$ and $j$ is defined
as the minimum number of nodes traversed by a path that goes from one
node to the other.  The clustering coefficient $c_i$ \cite{watts98} of
the node $i$ is defined as the ratio between the number of edges $e_i$
in the sub-graph identified by its nearest neighbors and its maximum
possible value $k_i(k_i-1)/2$, corresponding to a complete sub-graph,
{\em i.e.}  $c_i=2e_i/k_i(k_i-1)$.  This magnitude quantifies the
tendency that two nodes connected to the same node are also connected
to each other. The clustering coefficient $c_i$ takes values of order
${\cal O}(1)$ for grid networks. On the other hand, for random graphs
\cite{er,bollobas}, which are constructed by connecting nodes at
random with a fixed probability $p$, the clustering coefficient is of
order ${\cal O}(N^{-1})$.  Finally, the betweenness $b_i$ of a node
$i$ is defined as the total number of minimum paths that pass through
that node. It gives an measure of the amount of traffic that goes
through a node, if the minimum path distance is considered as the
metric defining the optimal path between pairs of nodes. The average
values of these metrics over every node (or pair of nodes for
$d_{ij}$) in the AS, AS+, and IR maps is given in Table~\ref{tab:1}.

\begin{table}[b]
\begin{tabular}{lrrlllll}
Map & $N$ & $E$ & $\langle k\rangle$ & $\langle d\rangle$ & $\langle
c\rangle$ & $\langle 
b\rangle/N$ \\
\hline
IR & 228298 & 320105 & 2.80 & 9.51 & 0.03  & 4.14\\
AS & 11174 & 23367 & 4.18 & 3.62 & 0.22  & 3.61\\
AS+ & 11461 & 32711 & 5.71 & 3.56 & 0.24 & 3.56\\
\hline
\end{tabular}
\caption{Average metrics  of the AS, AS+, and IR maps. 
See text for the metrics' definitions.} 
\label{tab:1}
\end{table}

\begin{figure}
\centerline{\psfig{file=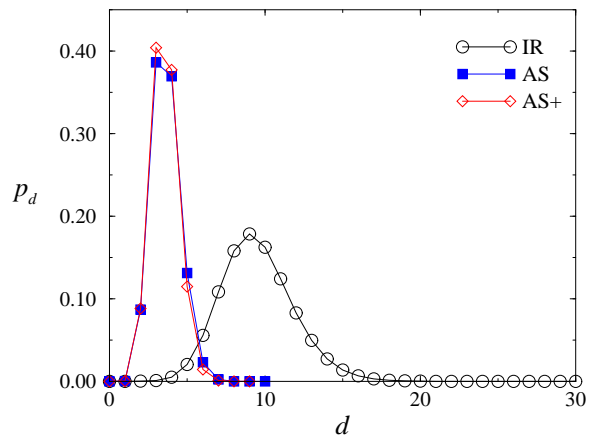,width=3in}}
\caption{Probability distribution $p_d={\rm
    Prob}[d_{ij}=d]$ of the minimum path distance between nodes, for
  the AS, AS+, and IR maps.}
\label{fig:1}
\end{figure}

The average connectivity for the three maps is of order ${\cal O}(1)$;
therefore, they can be considered as \textit{sparse} graphs.  Despite
the small average connectivity, however, the average minimum path
distance is also very small, compared to the size of the maps.  The
probability distribution of the minimum path distance, $p_d={\rm
  Prob}[d_{ij}=d]$, is shown in Fig.~\ref{fig:1}. For all maps this
distribution is sharply peaked around the average value $\left\langle d
\right>$; therefore, we can take $\left\langle d \right\rangle$ as the
characteristic minimum path distance. In the next section we will show
that this is not the case for the connectivity, that is characterized
by large fluctuations from node to node. Thus, the Internet strikingly
exhibits what is known as the ``small-world'' effect \cite{watts98}:
in average one can go from one node to any other in the system passing
through a very small number of intermediate nodes. Since the network
is sparse this necessarily implies that there are some hubs and
backbones which connect different regional networks, strongly
decreasing the value of $\left\langle d\right\rangle$.
The small world evidence is strenghtened by the empirical finding of
clustering coefficients for the AS, AS+, and IR  four orders
of magnitude larger than the corresponding value for a random graph of
the same size, ${\cal O} (N^{-1})$. As discussed above, this fact
implies that neighbors of the same node are very likely on their turn
connected among themselves. The high clustering coefficient of the
Internet maps is probably due to geographical constraint. In Internet
graphs, all links are equivalent.  Yet, the physical connections are
characterized by a real space length.  The larger is this length, the
higher the cost of installation and maintenance of the physical line,
favoring therefore the preferential connection between nearby nodes.
It is likely that nodes within the same geographical region will have
a large number of connections among them, increasing in this way the
clustering coefficient.

Another measure of interest is given by the number of minimal
paths that pass by each node. To go from one node in the network to
another following the minimum path, a sequence of nodes is visited.
If we do this for every pair of nodes in the network, there will be a
certain number of key nodes that will be visited more often than
others.  Such nodes will be of great importance for the transmission
of information along the network. This evidence can be quantitatively
measured by means of the betweenness $b_i$; {\em i.e.}  the number of
minimum paths that go through each node $i$. This magnitude has been
introduced in the analysis of social networks in Ref.~\cite{newman01}
and more recently it has been studied for the AS maps, with the name
of load \cite{goh01}.  An algorithm to compute the betweenness has
been described in Ref.~\cite{newman01}. For a star network the
betweenness takes its maximum value $N(N-1)/2$ at the central node and
its minimum value $N-1$ at the vertices of the star. The average
betweenness of the AS, AS+, and IR maps analyzed here is ${\cal
  O}(N)$, as shown in Table~\ref{tab:1}.  In the case of the AS and
AS+ maps, despite the enriched map has a much larger number of
edges, the average measures are very similar.

While some metrics are very alike (for instance, the average
betweenness $\left< b \right>$), some differences among others are
consistent with the fact that the AS and AS+ maps are a coarse-grained
representation of the IR map.  The IR level map is, for instance,
sparser, and its average minimum path distance is larger. The IR map
has a small average connectivity, because routers have a finite
capacity and, therefore, can have a limited number of connections. On
the contrary, ASs can have in principle any number of connections,
since they represent the aggregation of a large number of routers. 
This implies that AS maps have a greater number of nodes with a high number
of connections (hubs), providing the shortcuts needed to produce a
small average minimum path distance.

\section{Scale-free properties}

The analysis of the average measures presented in the previous section
makes clear that the Internet does not resemble a star-shaped
architectures with just a few gigantic hubs and a multitude of singly
connected nodes.  The same measurements rule out as well the
possibility of a random graph structure or a regular grid
architecture.  These evidences suggest a peculiar topology that will
be clearly identified by looking at the detailed distributions.  In
particular, Faloutsos {\em et al.} \cite{faloutsos} pointed out for
the first time that the connectivity properties of the Internet AS
maps are characterized by a probability distribution that a node has
$k$ links with the form $p_k \sim k^{-\gamma}$, where $\gamma\simeq 2.1$ is a
characteristic exponent.  This behavior signals the presence of
\textit{scale-free} connectivity properties; {\em i.e.} there is no
characteristic connectivity above which the probability is decaying
exponentially to zero. In other words, there is a statistically
significant probability that a node has a very large number of
connections compared to the average connectivity $\left <k \right>$.
In addition, the implicit divergence of $\left< k^2 \right>$ is
signalling the extreme heterogeneity of the connectivity pattern,
since it implies that statistical fluctuations are unbounded.  The
work of Faloutsos {\em et al.} was followed by different studies of AS
maps \cite{chou,us02}, AS+ maps \cite{chen02}, and IR maps
\cite{mercator,broido}.  Here, we will revisit the analysis of
scale-free properties in recent AS, AS+, and IR level maps.

\begin{figure}
\centerline{\psfig{file=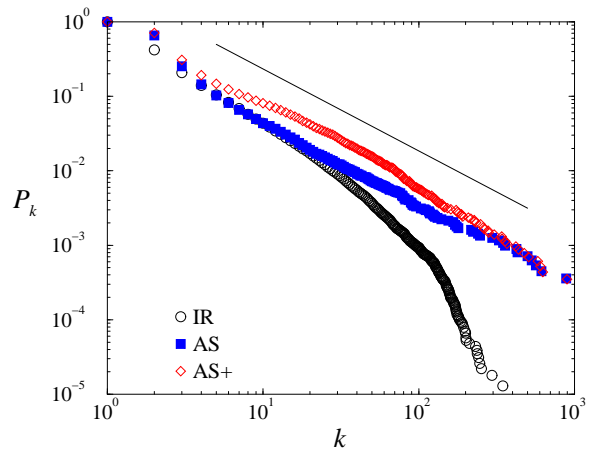,width=3in}}
\caption{Integrated connectivity  distribution   $P_k={\rm
    Prob}[k_i>k]$ for the AS, AS+, and IR maps.  The solid line
  corresponds to a power law decay $P_k\sim k^{1-\gamma}$ with exponent
  $\gamma=2.1$.}
\label{fig:2}
\end{figure}

We start by considering the integrated connectivity probability
$P_k={\rm Prob}[k_i>k]$. In the case of a pure power-law probability
distribution $p_k \sim k^{-\gamma}$, we expect the functional behavior
$P_k=ak^{1-\gamma}$, where $a$ is a normalization constant.  In
Fig.~\ref{fig:2} we show the connectivity distribution for the AS,
AS+, and IR maps. For the AS map a clear power law decay with exponent
$\gamma=2.1\pm0.1$ is observed, as it has been already reported
elsewhere~\cite{faloutsos,chou,us02}. The reported distribution is
also stable in time as found by analyzing different time snapshot of
the AS level maps obtained by the NLANR~\cite{us02}.  As noted in
Ref.~\cite{chen02}, the connectivity distribution for the AS+ enriched
data deviates from a pure power law at intermediate connectivities.
This anomaly might or might not be related to the biased enrichment of
the Internet sampling (see Ref.~\cite{chen02}).  While this represents
an important point in the detailed description of the connectivity
properties, it is not critical concerning the scale-free nature of the
Internet. With respect to the network physical properties, 
it is just the large connectivity region that is 
actually effective.  Indeed, recent studies
about network resilience to removal of nodes \cite{resilience} and
virus spreading \cite{spreading} have shown that the relevant
parameter is the ratio $\kappa=\left\langle k^2 
\right\rangle/\left\langle k \right\rangle$
between the first two moments of the connectivity distribution. If
$\kappa\gg1$ then the network manifests some properties that are not
observed for networks with exponentially bounded connectivity
distributions. For instance, we can randomly remove practically all
the nodes in the network and a giant connected component
\cite{bollobas} will still exist.  In both the AS and AS+ maps, in
fact, we observe a wide connectivity distribution, with the same
dependency for very large $k$. The factor $\kappa$ is mainly determined by
the tail of the distribution, and is very similar for both maps.  In
particular, we estimate $\kappa=265$ and $\kappa=222$ for the AS and AS+
maps, respectively.  With such a large values, for all practical
purposes (resilience, virus spreading, traffic, etc.)  the AS and AS+
maps behave almost identically.

\begin{figure}
  \centerline{\psfig{file=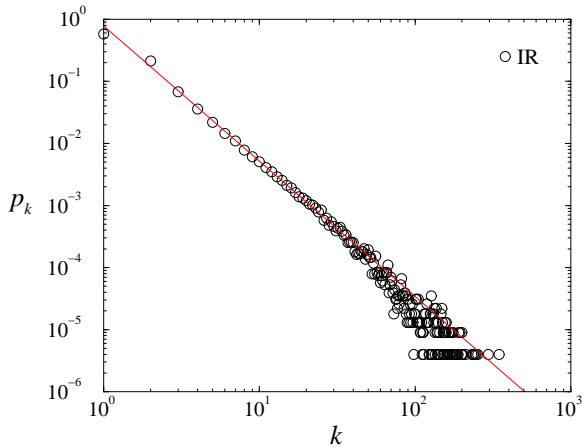,width=3in}}
  \caption{Connectivity distribution $p_k={\rm Prob}[k_i = k]$ for the
    IR map. The solid line is a power law decay $p_k\sim k^{-\gamma}$ with
    $\gamma=2.1$.}
\label{fig:3}
\end{figure}

The connectivity distribution of the IR level map has a power-law
behavior that is, however, smoothed by a clear exponential cut-off.
The existence of a power-law tendency for small connectivities is
better seen for the probability distribution $p_k={\rm Prob}[k_i=k]$,
as shown in Fig.~\ref{fig:3}.  A power law fit of the form
$p_k=a(1-\gamma)k^{-\gamma}$ for $k\leq300$ yields the exponent $\gamma=2.1\pm0.1$, in
perfect agreement with the exponent found for the integrated
connectivity distribution in the AS map. Nevertheless, for $k\gg 50$
the IR map integrated connectivity distribution follows a faster
decay.  This picture is consistent with a finite size scaling of the
form $p_k=k^{-\gamma}f(k/k_c)$ \cite{dmreview}. Here $k_c$ is a
characteristic connectivity beyond which the distribution decays
faster than a power law, and $f(x)$ has the asymptotic behavior
$f(x)=$ const. for $x\ll1$ and $f(x)\ll1$ for $x\gg1$. Deviations from the
power law behavior at large connectivities have been also observed for
the larger maps reported in Ref.~\cite{broido}. In that work, the
integrated probability distribution is fitted to the Weibull
distribution $P_k=a\exp[-(k/k_c)^\beta]$. While we do not want 
to enter into the details of the different fitting procedures, 
we suggest that the more general fitting form
$p_k=k^{-\gamma}f(k/k_c)$, in which $\gamma$ is an independent fitting
parameter, is likely a better option.

The presence of truncated power laws must not be considered a
surprise, since it finds a natural place in the context of scale-free
phenomena.  Actually, bounded scale-free distributions ({\em i.e.}
power-law distributions with a cut-off) are implicitly present in
every real world system because of finite-size effects or physical
constraints.  Truncated power laws are observed also in other real
networks \cite{amaral} and different mechanisms have been proposed to
explain the cut-off for large connectivities.  Actually, we can
distinguish two different kinds of cut-offs in real networks.  The
first is an exponential cut-off, $f(x) = \exp(-x)$, which can be
explained in terms of a finite connectivity capacity of the network
elements \cite{amaral} or incomplete information \cite{mossa02}.  This
is likely what is happening at the IR level, where the finite capacity
constraint (maximum number of router interfaces) is, in our opinion,
the dominant mechanism affecting the tail of the connectivity
distribution.  In this perspective, larger and more recent samples at
the IR level could present a shift in the cut-off due to the improved
technical router capabilities and the larger statistical sampling. A
second possibility is given by a very steep cut-off such as $f(x)
=\theta(1-x)$, where $\theta(x)$ is the Heaviside step function.  This is what
happens in growing networks with a finite number of elements. Since SF
networks are often dynamically growing networks, this case represents
a network which has grown up to a finite number of nodes $N$. The
maximum connectivity $k_c$ of any node is related to the network age.
The scale-free behavior is evident up the $k_c$ and then decays as a
step function since the network does not possess any node with
connectivity $k$ larger than $k_c$.  By inspecting Fig.~\ref{fig:2},
this second possibility appears realized at the AS level. Indeed, the
dominant mechanism at this level is the finite size of the network,
while connectivity limits are not present, since each AS is a
collection of a large number of routers, and it can handle a very
large connectivity load.

\begin{figure}
\centerline{\psfig{file=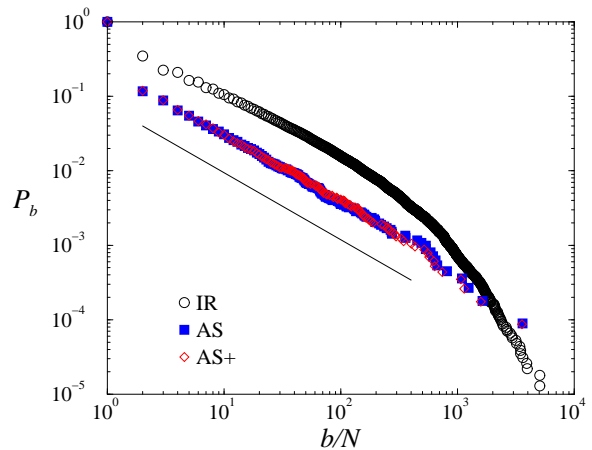,width=3in}}
\caption{Integrated betweenness distribution $P_b={\rm Prob}[b_i>b]$
  for the AS, AS+, and IR maps.  The solid line is a power law decay
  $P_b\sim b^{1-\gamma_b}$ with $\gamma_b=1.9$.}
\label{fig:4}
\end{figure}

The connection between finite capacity and bounded distributions
becomes evident also if we consider the betweenness. This magnitude is
a static estimate of the amount of traffic that a node supports.
Hence, if a router has a bounded capacity, the betweenness
distribution should also be bounded at large betweenness. On the
contrary, this effect should be absent for the AS maps. The integrated
betweenness distribution $P_b={\rm Prob}[b_i>b]$ for the AS, AS+, and
IR maps is shown in Fig.~\ref{fig:4}.  The AS and AS+ distributions
are practically the same and they are well fitted by a power law
$P_b\sim b^{1-\gamma_b}$ with an exponent $\gamma_b=1.9\pm0.1$. 
In the case of the
IR map, on the other hand, the betweenness distribution follows a
truncated power law, in analogy to what is observed for the
connectivity distribution. The betweenness distribution, therefore,
corroborates the equivalence between the AS and AS+ maps, and the
existence of truncated power laws for the IR map.

Finally, it is worth to stress that while the power law truncation is
an expected feature of finite systems, the scale-free regime is the
important signature of an emergent cooperative behavior in the
Internet dynamical evolution. This dynamics play therefore a central
role in the understanding and modeling of the Internet. In this
persepective, the developing of a statistical mechanics approach to
complex networks \cite{albert02} is providing a new dynamical
framework where the distinctive statistical regularities of the
Internet can be understood in term of the basic processes ruling the
appearance or disappearence of nodes and links.

\section{Hierarchy and correlations}

The topological metrics analyzed so far give us a distinction between
the AS and IR maps with respect to the large connectivity and
betweenness properties. The difference becomes, however, more evident
if we consider properties related with the existence of hierarchy and
correlations.  The primary known structural difference in the Internet
is the distinction between {\em stub} and {\em transit} domains.
Nodes in stub domains have links that go only through the domain
itself. Stub domains, on the other hand, are connected via a gateway
node to transit domains that, on the contrary, are fairly well
interconnected via many paths. This hierarchy can be schematically
divided in international connections, national backbones, regional
networks, and local area networks. Nodes providing access to
international connections or national backbones are of course on top
level of this hierarchy, since they make possible the communication
between regional and local area networks.  Moreover, in this way, a
small average minimum path length can be achieved with a small average
connectivity. This hierarchical structure will introduce some
correlations in the network structure, and it is an important issue to
understand how these features manifest at the topological level. In
order to exploit the presence of hierarchies in Internet
maps we introduce two metrics based on the clustering coefficient and
the nearest neighbor average connectivity \cite{us02}.

\begin{figure}
  \centerline{\psfig{file=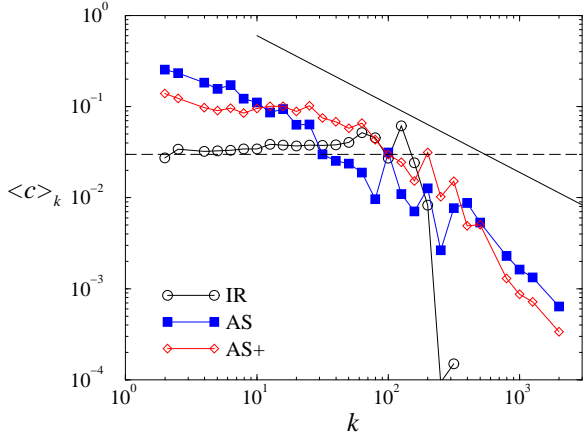,width=3in}}
  \caption{Average clustering coefficient as a function of 
    the node connectivity for the AS, AS+, and IR maps. The solid line
    is given by the power law decay $\left\langle c \right\rangle_k\sim
    k^{-0.75}$. The horizontal dashed line marks the average
    clustering coefficient $\left< c \right> = 0.03$ computed for the
    IR map.}
  \label{fig:5}
\end{figure}

The previously defined clustering coefficient is the 
average probability that two
neighbors $l$ and $m$ of a node $i$ are connected. Let us consider the
{\em adjacency matrix} $a_{ij}$, that indicates whether there is a
connection between the nodes $i$ and $j$ ($a_{ij}=1$), or the
connection is absent ($a_{ij}=0$). Given the definition of the
clustering coefficient, it is easy to see that the number of edges in
the subgraph identified by the nearest neighbors of the node $i$ can be
computed as $e_i = (1/2) \sum_{lm} a_{il}a_{lm}a_{mi}$. Therefore, the
clustering coefficient $c_i$ measures the existence of
\textit{correlations} in the adjacency matrix, weighted by the
corresponding node connectivity.  In section~\ref{sec:ave} we have
shown that the clustering coefficient for the AS, AS+, and IR maps is
four orders of magnitude larger than the one expected for a random
graph and, therefore, that they are far from being random. Further
information can be extracted if one computes the clustering
coefficient as a function of the node connectivity \cite{us02}. In
Fig.~\ref{fig:5} we plot the average clustering coefficient $\left\langle c
\right\rangle_k$ for nodes with connectivity $k$. In the case of the AS and
AS+ maps this quantity follows a similar trend that can be
approximated by a power law decay with an exponent around $0.75$.  For
the IR map, however, except for a sharp drop for large values of $k$,
attributable to low statistics, it is almost constant, and equal to
the average clustering coefficient  $\left< c \right> = 0.03$.
This implies that, in the AS and AS+ maps, nodes with a small number
of connections have larger local clustering coefficients than those
with a large connectivity.  This behavior is consistent with the
picture described in the previous section of highly clustered regional
networks sparsely interconnected by national backbones and
international connections. The regional clusters of ASs are probably
formed by a large number of nodes with small connectivity but large
clustering coefficients.  Moreover, they should also contain nodes
with large connectivities that are connected with the other regional
clusters. These large connectivity nodes will be on their turn
connected to nodes in different clusters which are not interconnected
and, therefore, will have a small local clustering coefficient.  On
the contrary, in the IR level map these correlations are absent.
Somehow the domain hierarchy does not produce any signature at the
single router scale, where the geographic constraints and connectivity
bounds probably play a more important role.

\begin{figure}
\centerline{\psfig{file=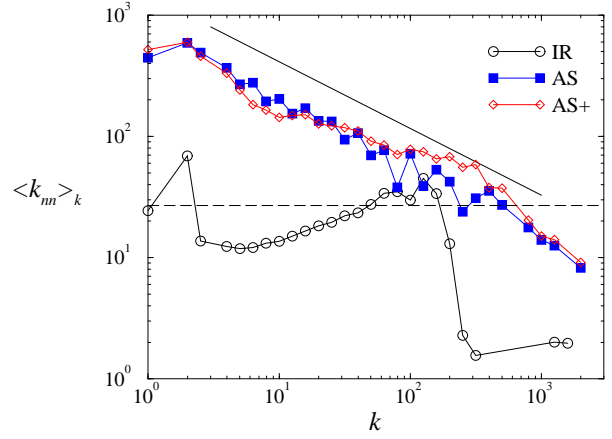,width=3in}}
\caption{Nearest neighbors average connectivity for the AS, AS+, and
  IR maps. The
  solid line is given by the power law decay $\langle k_{nn}\rangle_k\sim
  k^{-0.55}$. The horizontal dashed line marks the value in the
  absence of correlations, $\langle k_{nn}\rangle_k^0 = \left< k^2 \right> /
  \left< k \right> = 26.9$, computed for the IR map.}
\label{fig:6}
\end{figure}

These observations for the clustering coefficient are supported by
another metric related with the correlations between node
connectivities. These correlations are quantified by the probability
$p_c(q \,\vert\, k)$ that, given a node with connectivity $k$, it is
connected to a node with connectivity $q$. With the available data, a
direct plot of $p_c(q \,\vert\, k)$ results very noisy and difficult
to interpret~\cite{goh02}.  Thus in Ref.~\cite{us02} we suggested to
measure instead the nearest neighbors average connectivity of the
nodes of connectivity $k$, 
$\left\langle k_{nn} \right \rangle_k=\sum_q q \, p_c(q
\,\vert\, k)$, and to plot it as a function of the connectivity $k$.
If there are no connectivity correlations ({\em i.e.} for a random
network), then $p_c^0(q \,\vert\, k) = q \, p_q / \left< k \right>$,
where $p_q$ is the connectivity distribution, and we obtain $\langle
k_{nn}\rangle_k^0 = \left< k^2 \right> / \left< k \right>$, which is
independent of $k$.  The corresponding plots for the AS, AS+, and IR
maps are shown in Fig.~\ref{fig:6}. For the AS and AS+ maps we observe
a power-law decay for more than two decades, with a characteristic
exponent $0.55$, clearly indicating the existence of correlations. On
the contrary, the IR map displays again an almost constant nearest
neighbors average connectivity, very similar to the expected value for
a random network with the same connectivity distribution, $\langle
k_{nn}\rangle_k^0 \simeq 30$. Again, the sharp drop for large $k$ can be
attributed to the low statistics for such large connectivities.
Therefore, also in this case the two levels of representation show
very different features.

It is worth remarking that the present analysis of the hierarchical
and correlation properties shows a very good consistency of results in
the case of the AS and AS+ maps. This points out a robustness of these
features that can thus be considered as general properties at the AS
level. On the other hand, the IR map shows a marked difference that
must be accounted for when developing topology generators.  In other
words, Internet protocols working at different representation levels
must be thought as working on different topologies.  Topology
generators as well must include these differences, depending on the
level at which we intend to model the Internet topology.

\section{Conclusions} 

The increasing availability of larger Internet maps and the
proliferation of growing networks models with scale-free features have
recently stimulated a more detailed statistical analysis aimed at the
identification of distinctive metrics and features for the Internet
topology.  At this respect, in the present work we have presented a
detailed statistical analysis of several metrics on Internet maps
collected at the router and autonomous system levels.  Our analysis
confirms the presence of a power-law (scale-free) behavior for the
connectivity distribution, as well as for the betweenness
distribution, that can be associated to a measure of the load of the
nodes in the maps.  The exponential cut-offs observed in the IR maps,
associated to the limited capacity of the routers, are absent in the
AS level, which conglomerate a large number of routers and are thus
able to bear a larger load. The analysis of the clustering coefficient
and the nearest neighbors average connectivity show in a quantitative
way the presence of strong correlations in the Internet connectivity
at the AS level, correlations that can be related to the hierarchical
distribution of this network.  These correlations, on the other hand,
seem to be nonexistent at the IR level. The correlation properties
clearly indicate the presence of strong diferences between the IR and
AS levels of representation.  Our findings represent a step forward in
the characterization of the Internet topology, and will be helpful for
scrutinizing more thoroughly the actual validity of the network models
proposed so far, and as ingredient in the elaboration of new and more
realistic Internet topology generators. A first step in this direction
has been already given in the network model proposed in
Ref.~\cite{goh02}.

\section{Acknowledgments}

This work has been partially supported by 
the European Commission - Fet Open
project COSIN IST-2001-33555.  R.P.-S. acknowledges financial support
from the Ministerio de Ciencia y Tecnolog\'{\i}a (Spain).
We thank T. Erlebach for the help in the data collection process.

\end{document}